\documentclass[aps,prc,twocolumn,showpacs,preprintnumbers,amsmath,amssymb]{revtex4}
\usepackage{graphicx}
\usepackage{dcolumn}
\usepackage{bm}
\usepackage{comment}
\usepackage{rotating}
\usepackage{longtable}
\usepackage{float}
\usepackage{eucal}
\usepackage{csquotes}
\usepackage{xcolor} 
\makeatletter

\newcommand{\Rmnum}[1]{\expandafter\@slowromancap\romannumeral #1@}
\makeatother

\begin{document}

\title{\centering {Antimagnetic rotation and sudden change of electric quadrupole transition strength in $^{143}$Eu}}

\author{S. Rajbanshi$^{1}$}
\author{S. Roy$^{2}$}
\author{Somnath Nag$^{3}$}
\author{Abhijit Bisoi$^{1}$}
\author{S. Saha$^{2}$}
\author{J. Sethi$^{2}$}
\author{T. Trivedi$^{4}$}
\author{T. Bhattacharjee$^{5}$}
\author{S. Bhattacharyya$^{5}$}
\author{S. Chattopadhyay$^{1}$}
\author{G. Gangopadhyay$^{6}$}
\author{G. Mukherjee$^{5}$}
\author{R. Palit$^{2}$}
\author{R. Raut$^{7}$}
\author{M. Saha Sarkar$^{1}$}
\author{A. K. Singh$^{3}$}
\author{A. Goswami$^{1}$}
\email{asimananda.goswami@saha.ac.in}

\affiliation{$^1$Saha Institute of Nuclear Physics, Kolkata 700064, India}
\affiliation{$^2$Tata Institute of Fundamental Research, Mumbai 400005, India}
\affiliation{$^3$Indian Institute of Technology, Kharagpur 721302, India}
\affiliation{$^4$Guru Ghasidas Vishayavidyalaya, Bilaspur 495009, India}
\affiliation{$^5$Variable Energy Cyclotron Center, Kolkata 700064, India}
\affiliation{$^6$University of Calcutta, Kolkata 700009, India}
\affiliation{$^7$UGC-DAE consortium for scientific Research, Kolkata 700098,
 India}

\date{\today}

\begin{abstract}
Lifetimes of the states in the quadrupole structure in $^{143}$Eu have been 
measured using the Doppler shift attenuation method as well as parity of the
 states in the sequence has been firmly identified from polarization 
measurement using the Indian National Gamma Array. The decreasing trends of 
the deduced quadrupole transition strength $B(E2)$ with spin, along with 
increasing $\it{J}$$^{(2)}$/$B(E2)$ values before band crossing, conclusively
 establish the origin of these
 states as arising out of antimagnetic rotation. The abrupt
 increase in the $B(E2)$ values after the band crossing in the quadrupole band,
 a novel feature observed in the present experiment, may indicates the
 crossing of different shears configurations resulting in re-opening of shears
 structure. The results are well reproduced by numerical calculation within
 the framework of semi-classical geometric model.

\end{abstract}

\pacs{21.10.Re, 21.10.Tg, 21.60.Ev, 23.20.Lv, 27.60.+j}

\maketitle

It is well known that the deviation of a quantal system from spherical
 symmetry results in the observation of regular band like structure in its
 excitation spectrum, with the excitation energy proportional to $I(I+1)$,
 $I$ being the quantized angular momentum of a state \cite{abohr}. The
 deformed nuclei, found in the specific mass regions of the periodic table
 \cite{defor1, defor2}, are the best examples of such quantum rotation. In
 true sense it is the charge density that deviates from the spherical
 symmetry which specifies the orientation of the system as a whole. The
 resulting sequence of rotational levels gives rise to a band structure
 where the states are connected through strong E2 transitions. The
 experimental signature of these bands is the observation of almost
 constant electric quadrupole transition rates which are proportional to
 the square of the electric quadrupole moment operator \cite{abohr1}.

Interestingly, similar regular sequences of the quadrupole transitions have
 also been observed for nuclei with small quadrupole collectivity, but with
 different intrinsic properties compared to the strongly deformed systems
 \cite{cjchi, msuag, phreg}. These type of excitation mechanism was first
 reported by Zhu \textit{et al.} \cite{szhu} from the spectroscopic 
investigation of $^{100}$Pd nucleus. These bands are characterized by 
decreasing electric quadrupole transition rates [$B(E2)$] with increasing 
spin \cite{ajsim, pdatt, ajsim1, deep1, deep2, santo, nather}. The intriguing 
feature of the falling trend of the $B(E2)$ values of these sequences has 
been visualized as a new form of quantized rotation,
 known as antimagnetic rotation (AMR) that has also been interpreted in the
 framework of shears mechanism \cite{frauen1}. In this description the angular
 momentum is generated by closing of the two blades of conjugate shears,
 produced by the valence particles (holes). These valence particles (holes)
 are initially aligned in the time reversed orbit at the band head
 \cite{frauen1}. There is no net perpendicular component of the magnetic
 dipole moment for this configuration and it is symmetric with respect to
 rotation by $\pi$ about the total angular momentum axis (rotational axis).
 The resulting quadrupole transition strength will decrease with the increase
 of spin along the band due to the gradual closing of the angular momentum
 blades.

Another type of regular band-like structure with different characteristic
 features but same decreasing trend of $B(E2)$ values was observed for
 several nuclei in the mass $A$ $\sim$ 110 and 160 regions \cite{afan}
 and has been interpreted as smoothly terminating bands. These bands show
 the characteristic of gradually decreasing dynamic moments of inertia with
 increasing spin in contrast to fairly constant dynamic moment of inertia
 (without any collective contribution) in the case of the antimagnetic
 rotation \cite{ajsim1}. The calculations show that these bands arise when
 a particular configuration evolves continuously from high collectivity at
 low spin to a point where all the spin vectors of the valance nucleons,
 are aligned. With increasing spin the intrinsic shape evolves until it is
 symmetric around the axis of rotation. Since collective rotation about the
 symmetry axis is forbidden, no further angular momentum can be generated
 through collective rotation and thus represents the termination of the
 rotational band. The difference between this mechanism and the antimagnetic
 rotation is reflected in the variation of dynamic moment of inertia
 [$\it{J}$$^{(2)}$] and $B(E2)$ strength as a function of spin. In case of
 smoothly terminating bands the ratio $\it{J}$$^{(2)}$/$B(E2)$ remain almost
 constant in contrast to sharp increase in the case of the antmagnetic
 rotation \cite{ajsim1}.

The observation of conjugate shear structure responsible for the generation
 of angular momentum in near spherical systems in the form of AMR is also
 associated with the possibility of a similar complementary excitation mode
 called magnetic rotation (MR) due to single shear structure \cite{frauen1}.
 Indeed different manifestation of the shears mechanism with single shear
 structure have been found in mass regions viz. $A$ $\sim$ 80, 100, 140 and
 190 \cite{gbald, pagar, slaks, jenkins, nskel, rschwe, rschwe1, pdatta,
 amita, hubel}. Since both these two types of quantized rotation are the
 consequence of the shear mechanism, it is expected to observe both of them
 in the mentioned mass regions. However, till day, simultaneous occurrence
 of these two phenomena have been observed only in mass $\sim$ 100 region
 where firm experimental evidence of antimagnetic rotation has been reported
 in several Cd \cite{ajsim, pdatt,ajsim1, deep1, deep2, santo} isotopes and
 in the $^{104}$Pd nucleus \cite{nather} along with the observation of MR
 bands \cite{nskel, pdatta}. These bands have been interpreted in the
 framework of simple geometric model  \cite{rmcla1, santo, srsc1} and as
 well as the fully self-consistent microscopic tilted axis cranking method
 based on covariant density functional theory \cite{pwzha}. For the mass 
$A \sim 140$ region, the observed quadrupole band in $^{144}$Dy \cite{msuga1} 
has been identified as a possible candidate for antimagnetic rotation on the 
basis of theoretical arguments. However, due to the absence of lifetime data, 
the nature of excitation mechanism for this band cannot be firmly establish 
as antimagnetic rotation.

In the present letter, observation of AMR band is reported for the
 $^{143}$Eu nucleus. The
 AMR phenomenon in the present case has been established on the basis of the
 decreasing $B(E2)$ values with increasing spin for the band of interest.
 Further-more, a sudden increase of the $B(E2)$ value, at a higher spin,
 followed by a another rapid decrease has been observed. This is the first
 nucleus to exhibit such a behaviour and novel feature in the context of
 the AMR band.

\begin{figure}
  \centering
\vskip -6mm
 \includegraphics[clip=true,width=.45\textwidth,height=10.0cm]{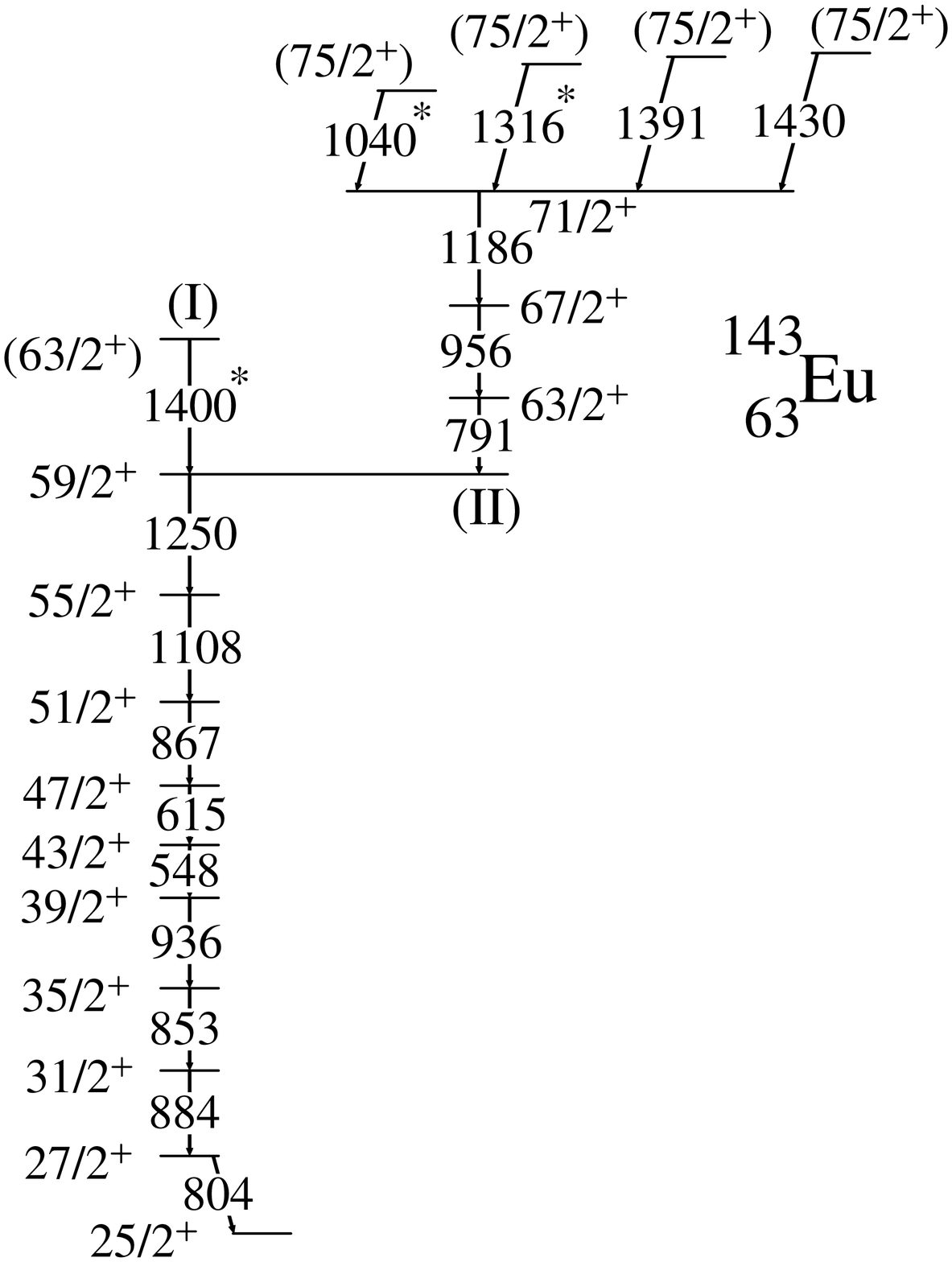}
\llap{\includegraphics[height=5.55cm]{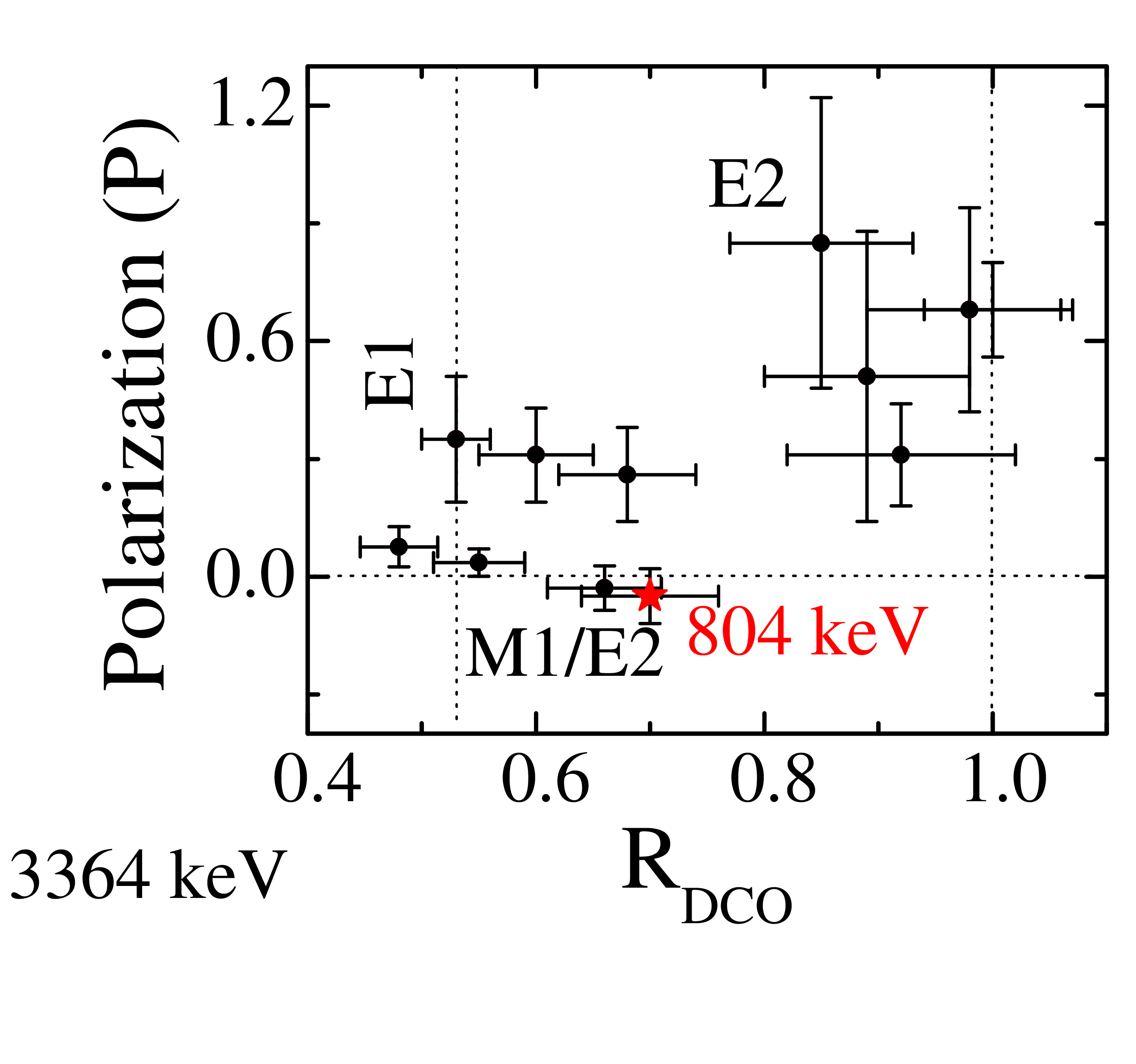}} \\

\vskip -2mm

\caption{\label{level} (Color  online) The partial level scheme of the
 quadrupole structure in $^{143}$Eu. In the inset figure shows the variation
 of polarization vs DCO ratios (gate on $E2$
 transitions) for different transitions in $^{143}$Eu. The value for the 803.9
 keV is represented by the filled red star. The level energy and gamma
 energies are rounded off to the nearest keV. The new transitions are indicated
 by an asterisk.}
\end{figure}

The lifetimes of the excited states in $^{143}$Eu have been measured using
 Doppler shift attenuation method (DSAM). The excited states in the $^{143}$Eu
 nucleus have been populated through the $^{116}$Cd ($^{31}$P, 4n) reaction
 at a beam energy of 148-MeV. The beam was delivered by the Pelletron
 Linac Facility at the Tata Institute of Fundamental Research (TIFR), Mumbai.
 The target was 2.4 mg/cm$^{2}$ thick $^{116}$Cd, enriched to 99\%, on 14.5
 mg/cm$^{2}$ Pb backing. Indian National Gamma Array (INGA)
 \cite{palit, palit1}, consisting of nineteen Compton suppressed clover
 detectors, was used to detect the de-exciting ${\gamma}$ rays. The
 experimental aspects and data analysis are detailed in Ref. \cite{rajban}.

\begin{figure}
\centering
\includegraphics[width=0.45\textwidth]{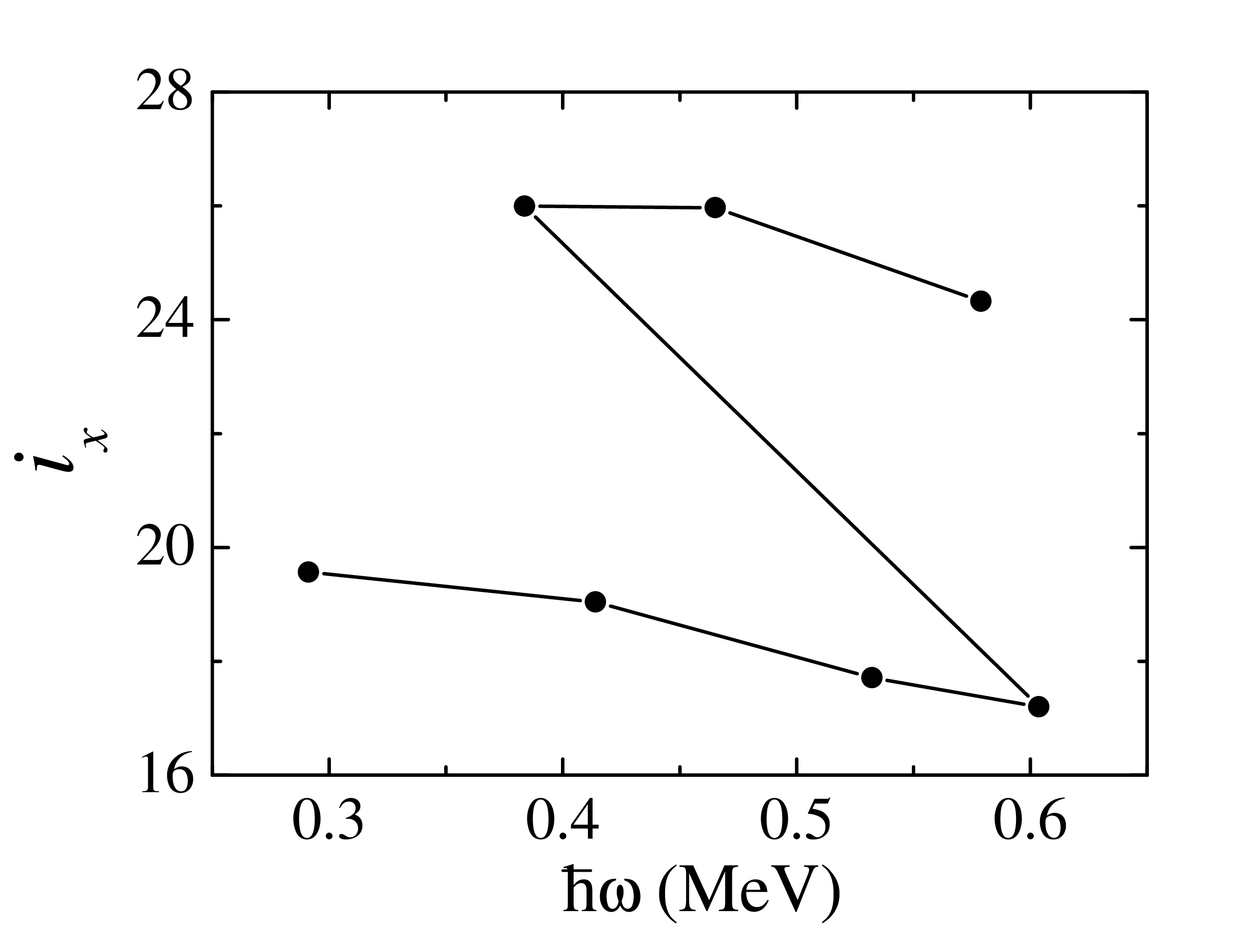}
\vskip -3mm
\caption{\label{aling} Experimental aligned angular momentum ($\it{i}$$_{x}$)
 for the quadrupole structure (above 43/2$^{+}$) in $^{143}$Eu. The reference
 parameters were assumed to be ${\mathcal{J}_{0}}$ = 12 $\hbar^{2}$$MeV^{-1}$
 and ${\mathcal{J}_{1}}$ = 25 $\hbar^{4}$$MeV^{-3}$,
 adopted from ref. \cite{cullen}.}
\end{figure}

The earlier studies on $^{143}$Eu \cite{piipar} reported a quadrupole cascade
 of $E2$ transitions, connected to the lower part of the level scheme by a
 803.9-keV $E1$ transition. All the reported transitions \cite{piipar} have 
been observed in the present measurements except for three new transitions 
which were placed in the partial level scheme [Fig. \ref{level}]. The values
 of DCO ratio, anisotropy and linear 
polarization determined from the present experiment for the 803.9 keV 
transition have been found to be 0.70(0.06), 0.78(0.09) and -0.05(0.07), 
respectively [inset of Fig. \ref{level}]. These values conclusively establish
 that the 803.9 keV transition is of mixed character [with $M1/E2$ mixing
 ratio 0.15(0.05)]. Thus the reported $E2$ cascade has been identified as a
  positive parity sequence [Fig. \ref{level}]. In addition to the previously
 observed transitions, a weak 1400.0-keV
 transition parallel to 790.9-keV $\gamma$ ray is observed above the 1249.7-keV
 transition. We have also observed two more weak transitions of energies
 1316.0 and 1040.0 keV feeding the 71/2$^{+}$ state. Fig. \ref{aling} shows the
 alignment plot for the quadrupole structure in $^{143}$Eu which is  indicative
 of a band crossing at a spin of 59/2$^{+}$.

\begin{table*}
\centering
\caption{\label{table1} Measured level lifetimes ($\tau$) and the corresponding $B(E2)$ values for the
 quadrupole transitions in $^{143}$Eu. Dynamic moment of inertia
 $\it{J}$$^{(2)}$ and the ratio of $\it{J}$$^{(2)}$/$B(E2)$ are also shown
 for the same transitions.}
\begin{ruledtabular}
\begin{tabular}{cccccccc}

Sequence & $I_{i}^{\pi}$  & $E_\gamma$   & ${\tau}$$^{a}$ & $\tau$$^{b}$ & $B(E2)$  & $\it{J}$$^{(2)}$   & $\it{J}$$^{(2)}$/$B(E2)$  \\
         &  [$\hbar$]     &    [keV]     & [ps] &
[ps]     &  [e$^{2}$b$^{2}$] & [$\hbar$$^{2}$MeV$^{-1}$] & 
[$\hbar$$^{2}$MeV$^{-1}$/(eb)$^{2}$] \\

\hline\hline

    & 47/2$^{+}$ & 614.8  & 3.1(3) & 2.76$_{-0.35}^{+0.42}$ & 0.34$_{-0.04}^{+0.05}$ 
& 15.9 & 47$^{+8}_{-7}$ \\
Structure I &51/2$^{+}$ & 866.5  & 0.6(2) & 0.66$_{-0.10}^{+0.14}$ &
 0.25$_{-0.04}^{+0.05}$  & 16.6 & 66$^{+13}_{-11}$ \\
    &55/2$^{+}$ & 1107.5 & $<$ 0.5 & 0.37$_{-0.06}^{+0.08}$ & 0.13$_{-0.02}^{+0.03}$
 & 28.1 & 216$^{+50}_{-33}$ \\
    &59/2$^{+}$ & 1249.7 & $<$ 0.5 & 0.44$_{-0.05}^{+0.06}$ & 0.06$_{-0.01}^{+0.01}$
 & 26.6 & 444$^{+74}_{-74}$ \\
\\
    &63/2$^{+}$ & 790.9 & $<$ 0.5 & 0.48$_{-0.07}^{+0.08}$ & 
  0.35$_{-0.05}^{+0.06}$$^{d}$ & 24.3 & 69$^{+12}_{-10}$ \\

Structure II &67/2$^{+}$  & 955.6 & 0.22(5)  & 0.44$_{-0.05}^{+0.07}$ & 
0.23$_{-0.03}^{+0.04}$  & 17.4 & 76$^{+13}_{-10}$ \\

    &71/2$^{+}$ & 1185.7 & $<$ 0.3 & 0.38$^{c}$ &  0.09$\uparrow$ 
& 19.5 & 216$\downarrow$ \\

\hline\hline

\multicolumn{6}{l}{$^{a}$The level lifetimes are adopted from ref \cite{piipar}.}\\

\multicolumn{6}{l}{$^{b}$Present measurements.}\\

\multicolumn{6}{l}{$^{c}$Upper limit of level lifetime (${\tau}$).}\\

\multicolumn{6}{l}{$^{d}$The 65\% branching ratio for the 790.9 keV
 transition \cite{piipar}.}\\

\end{tabular}
\end{ruledtabular}
\end{table*}

The Doppler-broadened line shapes have been observed for the transitions above
 the $I^{\pi}$ = 43/2$^{+}$ state in $^{143}$Eu. The level lifetimes of the states
 have been extracted by fitting these line shapes using the LINESHAPE analysis
 code \cite{wells, johnson}. The slowing down history of 10,000 recoil nuclei,
 traversing the target and the backing media, have been simulated by Monte
 Carlo techniques with a time step of 1.5 fs. The Shell corrected stopping
 powers of Northcliffe and Schilling \cite{north} have been used for these
 calculations. These histories have been used to generate angle dependent
 velocity profiles for detectors at different angles wherein the clover geometry
 of the detector has been used as an input. The velocity profiles have been
 used to calculate Doppler shapes for the $\gamma$-ray transitions of
 interest. The experimental spectra have been constructed with gates on the
 $\gamma$-ray transitions below the band of interest. The lifetimes for the
 states in the band have been extracted by least square fitting of the
 calculated shapes to the experimental (gated) spectra. The gate on the
 transitions below the transitions of interest necessitates to consider
 the side-feeding contribution. This has been modeled with a cascade
 of five transitions and having a moment of inertia same as that of the
 band under consideration. Initially, starting from the topmost transition,
 the members of the band have been sequentially fitted. A direct feeding has
 been assumed to calculate the shape for the transition from topmost level
 ($I^{\pi}$ = 71/2$^{+}$) for which a clear line shape has been observed. This gives us the value of
 the effective lifetime of the state. For the subsequent transitions in the
 band, the transition quadrupole moment, the side-feeding quadrupole moment,
 the peak height and the background have been used as the free parameters for
 the least square procedure. Following a satisfactory fit, the spectrum
 parameters like the peak height and the background have been fixed at the
 corresponding values. After having fitted all the transitions of the band,
 sequentially, a global least square minimization has been carried out for
 all the transitions of the cascade, simultaneously, wherein only the
 transition quadrupole moment and the side-feeding quadrupole moments for
 each state have been kept as free parameters.

\begin{figure*}
 \centering

  \includegraphics[width=.70\textwidth]{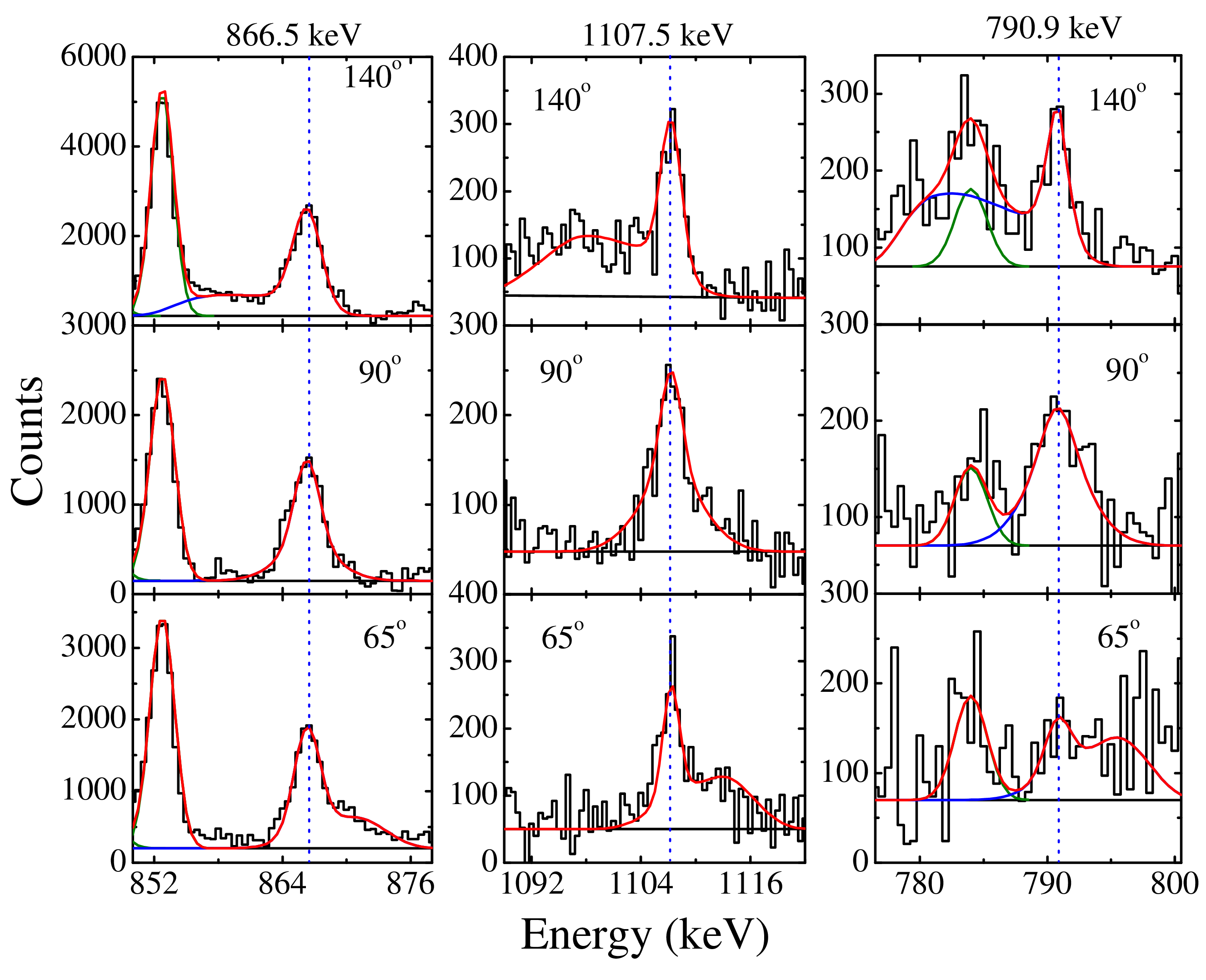} \\

\caption{\label{shapequ} (Color  online) The representative spectra along
 with theoretically fitted line-shapes for the quadrupole transitions 866.5,
 1107.5 and 790.9 keV of quadrupole structure in $^{143}$Eu. Calculated
 line-shape of ${\gamma}$ transitions, contaminant peaks and total line-shapes
 are represented by the blue, olive and red curves, respectively.}
\end{figure*}

In the present work, the observed line shapes at angles 65$^{\circ}$, 
90$^{\circ}$ and 140$^{\circ}$ have been fitted simultaneously to obtain the
 level lifetimes recorded in Table \ref{table1} along with the derived $B(E2)$ values. 
The representative line shape fits are illustrated in Fig. \ref{shapequ}.
 The uncertainties in the lifetimes have been derived from the behavior of the
 $\chi^{2}$ fit in the vicinity of the minimum. The quoted errors in the
 lifetimes do not include the systematic error due to the uncertainty in the
 stopping power which can be as large as 15\%. The level lifetimes have been evaluated in the present analysis by considering the side feeding from both observed as well as unobserved transitions to the level under consideration \cite{pdatta}. Thus the lifetime for the 47/2$^{+}$ state has been evaluated by including the 969 keV transition \cite{piipar} in the feeding history parallel to the 866.5 keV transition. The 969 keV transition appears to be a fully stopped peak in the present experiment. For the next higher lying state at 51/2$^{+}$, it was observed that the state was populated by a 698 keV transition (53/2 $\rightarrow$ 51/2$^{+}$; not shown in Fig. \ref{level}) in addition to the 1107.5 keV transition also. We have observed line shapes for both these transitions in the experimental spectra. The partial lifetime for the 51/2$^{+}$ state due to the feeding of 698 keV transition has been evaluated. In the final analysis, the top feed lifetime for the 51/2$^{+}$ level was assumed to be the intensity weighted average of the lifetimes for 55/2$^{+}$ and 53/2 levels since this level was fed by both 1108 keV (55/2$^{+}$ $\rightarrow$ 51/2$^{+}$) and 698 keV (53/2 $\rightarrow$ 51/2$^{+}$) $\gamma$ rays \cite{niyaz1}. The side feeding intensity in all the levels was fixed to reproduce the observed intensity pattern at 90$^{\circ}$ with respect to the beam direction. The present analysis is
 validated by the close compliance of the lifetimes of 47/2$^{+}$ and 
51/2$^{+}$ states, measured in the present work, with those reported in Ref.
 \cite{piipar}. For the states above the 51/2$^{+}$ level, Ref. \cite{piipar}
 only provided an upper limit on the lifetimes except for the 67/2$^{+}$
 level.

\begin{figure}[t]
\centering
\includegraphics[width=0.38\textwidth]{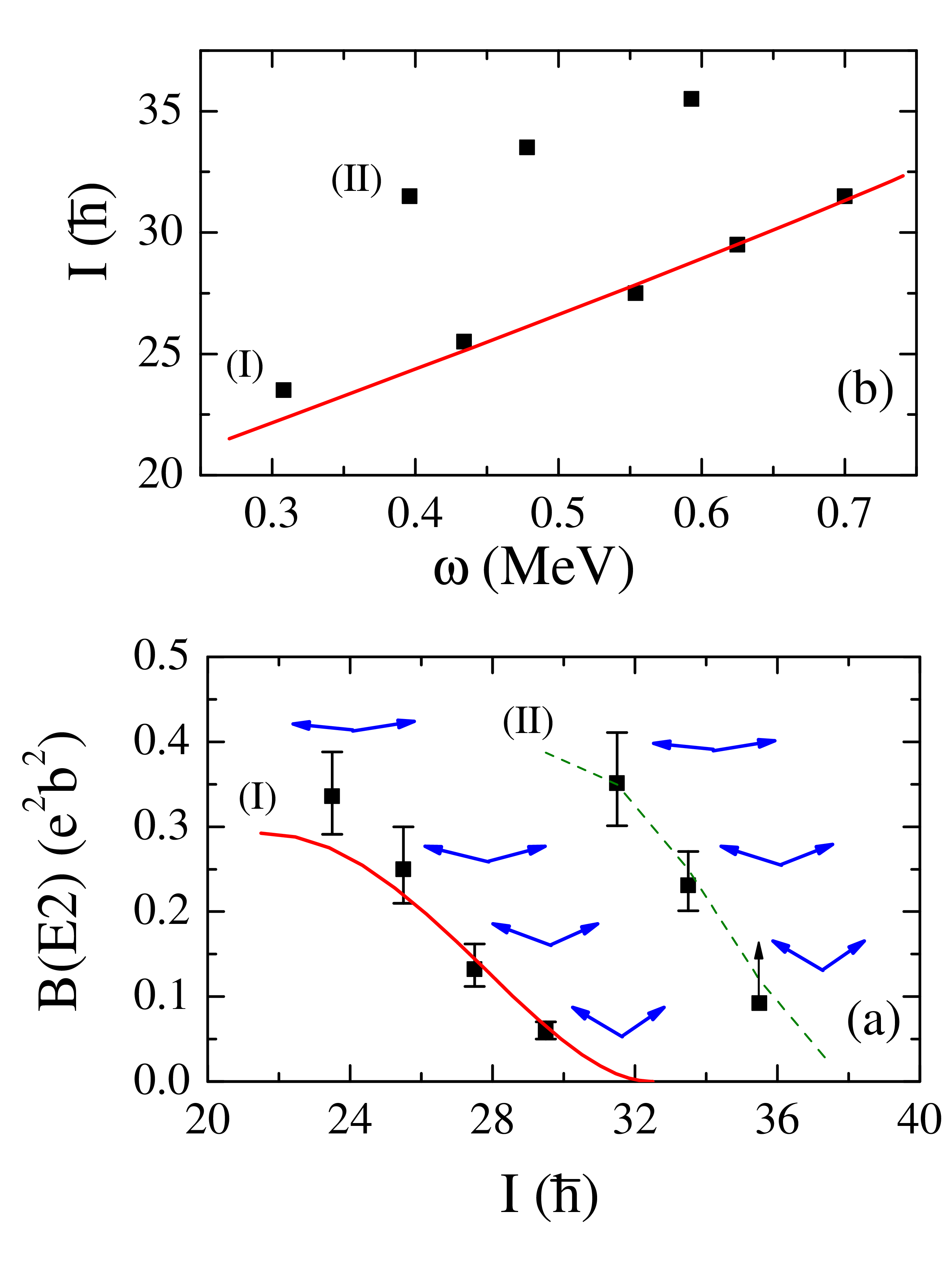}
\caption{\label{be2tve} (Color online) Experimental (a) $B(E2)$ values with
 spin [${I}$($\hbar$)] and (b) ${I}$($\hbar$) against rotational frequency
 ($\omega$) for quadrupole band in $^{143}$Eu. The solid red line represents
 the theoretical  semi-classical particle plus rotor model calculation. The
 parameters used for this calculation are $V_{\pi}$$_{\nu}$ = 1.7-MeV,
 $V_{\pi}$$_{\pi}$ = 0.2-MeV, $j$ = 11/2, $n$ = 10, $a$ =3.91 and $eQ_{eff}$
 = 1.40 eb. The parameters have same meaning as in Ref. \cite{santo, srsc1}.
 The arrows depict the relative orientation of the $h_{11/2}$ proton blades
 for structure I and structure II. Theoretical $B(E2)$ values, represented by olive dash line
 ($j_{h}$ = 59/2$\hbar$, $j_{\pi}$ = 9/2$\hbar$ and $eQ_{eff}$ = 1.58 eb),
 are calculated using Eq. (1) for structure II.}
\end{figure}

The Figure \ref{be2tve} (a) shows the variation of the deduced $B(E2)$ values
 with spin for the observed quadrupole band in $^{143}$Eu. The $B(E2)$ value shows
 a rapid decrease up to the state with spin-parity 59/2$^{+}$ (marked as (I)
 in Fig. \ref{be2tve} (a)). For the next higher lying state at 63/2$^{+}$ it
 shows a sudden increase and again continues to decrease along the band
 (marked as (II) in Fig. \ref{be2tve} (a)). The rapid increase of 
$\it{J}$$^{(2)}$/$B(E2)$ ratios with increasing spins (Table \ref{table1}) before band
 crossing clearly exclude the possibility of the structure I having a smoothly
 terminating origin. The trend of the $B(E2)$ values and 
$\it{J}$$^{(2)}$/$B(E2)$ ratios are the definitive experimental signature for
 the AMR phenomenon so far as structure I in concerned. This is the conclusive experimental evidence
 of the AMR band observed in any mass region other than $A$ ${\sim}$ 100. The behaviour of the
 structure II will be discussed later.

The $^{143}$Eu ($Z = 63, N = 80$) nucleus has one proton hole and two neutron
 holes with respect to the semi-magic nucleus $^{146}$Gd. However protons can
 be easily excited to the $\it{h}$$_{11/2}$ orbital across $Z = 64$ subshell
 closure, leading to the observation of the magnetic rotational bands. Such
 bands have been observed in the neighboring $^{142}$Sm, $^{141}$Eu,
 $^{142}$Gd nuclei, which were interpreted in the framework of the tilted axis
 cranking and the shears mechanism with the principal axis cranking model using
 small oblate deformation \cite{rajban, podsvi, pasern}.

In contrast to the previous work \cite{piipar}, we have assigned the configuration ($\nu{h}_{11/2}^{-2}$ $\pi({d}_{5/2}/{g}_{7/2})^{-3}$)$_{37/2}$
 ${\otimes}$ ($ \pi h_{11/2}^{2}$)$_{0}$ to the 43/2$^{+}$ state (previous assignment was $\pi({d}_{5/2}^{-1}{g}_{7/2}^{-1} h_{11/2})$ $\nu{h}_{11/2}^{-2}$). This is due 
to the fact that for oblate deformation, three proton holes in the 
$d$$_{5/2}$/$g$$_{7/2}$ orbitals, being rotation aligned, along with two
 neutron holes in the $h$$_{11/2}$, produce a total angular momentum of
 37/2$^{+}$. This is close to the band head spin of 43/2$^{+}$ for the proposed
 antimagnetic rotational band (structure I). The difference (3$\hbar$) can be
 attributed to the contribution of the core. On the other hand, the two protons
 in the time reversed $h$$_{11/2}$ orbital produce two angular
 momentum vectors which are anti-aligned to each other and perpendicular to the
 total angular momentum of the rotation aligned holes. Thus, for this 
configuration the double shear structure can exist. We have calculated the
 band head energy for the above configuration
  using a particle hole calculation in the relativistic mean field approach
 \cite{rraut} using the blocked BCS method.  The calculated energy of the band
 head is found to be 6.8 MeV whereas the corresponding observed value is 7.4
 MeV. The difference may be due to the effect of core rotation which is not
 considered in the present relativistic mean field calculations.  It has also
 been observed from  the calculations that  the potential energy surfaces for
 the seven quasi-particle state is much deeper ($\sim$ 700 keV) for oblate
 deformation compared to prolate deformation.

In order to explore the possibility of the antimagnetic rotational band in $^{143}$Eu for the above mentioned configuration, we
 have performed a numerical calculation within the framework of a self-consistent semi-classical rotor plus shears model \cite{santo, srsc1} based on the proton-neutron residual interaction \cite{aomac}. In this model the total energy of an excited state is
 the sum of the rotational energy of the weakly deformed core and the
 effective interaction energy between the shears blades. The transition
 probability of the state in the antimagnetic shear can be expressed as
 \cite{santo, srsc1},

\begin{equation}
B(E2)=\frac{15}{32\pi}(eQ_{eff})^{2} sin^{4}\theta
\end{equation}

where $\theta$ is the angle between the rotational axis and any one
 of the proton angular momentum vectors.

Experimental $\it{I}$($\omega$) values have been compared with the
 calculations from the semi-classical rotor plus shears model in Fig.
 \ref{be2tve}(b). The good agreement seems to indicate that the structure I originates
 due to the antimagnetic rotation with the configuration
 $\nu{h}_{11/2}^{-2}$ $\pi({d}_{5/2}/{g}_{7/2})^{-3}$ ${\otimes}$
 $\pi h_{11/2}^{2}$. In order to validate this proposition, the 
 $B(E2)$ values have been computed using Eq. (1) where the shears angle for
 each angular momentum state has been calculated from the semi-classical
 model. These values are represented by the solid line in Fig. \ref{be2tve}(a).
 This agreement provides the essential self-consistency
 check. The presence of the 1400.0-keV
 transition above the 59/2$^{+}$ state may indicate the termination of
 the aforesaid band at a spin of 63/2$^{+}$ state due to the complete alignment
 of two $h_{11/2}$ protons leading to an angular momentum gain of 10$\hbar$.
 Thus, the present calculations seems to indicate that the sequence I
 originates due to the AMR in a oblate nucleus. This is a unique observation
 since all the nuclei of the $A$ $\sim$ 100 are prolate where the AMR has
 been established.

It has been observed from Fig. \ref{be2tve} (a) that there is a large increase
 of $B(E2)$ value at a spin of 63/2$^{+}$. This nature of variation in the
 $B(E2)$ value, an unique feature of the current observation, has not observed
 in any nucleus. However, such sudden increase in $B(M1)$ values observed in
 different mass regions and have been interpreted as a crossing of the two
 magnetic rotational bands \cite{clark, pagar, santo2}. Thus, the observation 
may be associated to the emergence of a new double shear structure. This 
possibility is also supported by the fact that the original AMR structure can
 generate spin only upto 63/2$^{+}$ and the observed states with higher 
angular momentum must have a different single particle configuration, 
resulting in a re-opening of the shear structure. A new configuration
 [$\nu{h}_{11/2}^{-2}$ $\pi({d}_{5/2}/{g}_{7/2})^{-3}\pi h_{11/2}^{2}$ + 
$core$(3$\hbar$)]$_{59/2}$ ${\otimes}$ $ \pi h_{11/2}^{2}$ has been tentatively
 assigned for the quadrupole structure II where two additional $h_{11/2}$ 
protons in the time reversed orbit start to align. This is in addition to the 
rotation aligned spin (59/2) due to the structure I. The coupling of these 
angular momentum vectors may be visualize as another twin shear like structure
 responsible for the generation of a different AMR band with a new shear angle.

The semi-classical model calculation agreed well with the experimental $B(E2)$
 values for the structure II (Fig. \ref{be2tve}(a)) assuming the fact that
 the proton blades are not fully stretched. Such non-stretched configurations have been
 assumed in semi-classical calculation \cite{podsvi, pasern, rajban} in this
 mass region. In this calculation, the values of shears angle ($\theta$) are
 obtained directly from the relation $\it{I}$ = $\it{j}$$_{h}$ + 2$\it{j}$$_{\pi}$cos$\theta$ for the assumed
 configuration where $j_{h}$ is the total angular momentum of the quasi 
particles aligned along the rotational axis at the band head and the angular
 momentum assumed due to the rotation of the core. In this case the maximum 
spin that can be generated due to the alignment of two more protons
 in $h_{11/2}$ is 6$\hbar$ because of Pauli
 blocking (as two protons are already aligned corresponding to the projection
 11/2 and 9/2 for the sequence I, the available projected states for the next
 pair of $h_{11/2}$ protons are 7/2 and 5/2) and hence, this band is expected to generate a maximum spin of 71/2$^{+}$ which corroborated with the experimental results. The presence of high energy transitions above the state at 71/2$^{+}$ may visualize as the termination of the AMR band. However, the increasing trend of the $\it{J}$$^{(2)}$/$B(E2)$ ratios for the structure II [Table \ref{table1}] can not be definitely ascertain from the present experimental data. Hence the possibility of smooth band termination for  structure II cannot be excluded \cite{afan}.

In summary, the high spin quadrupole structure in $^{143}$Eu has been
 investigated using the reaction $^{116}$Cd ($^{31}$P, 4n) at a beam energy
 of 148-MeV. Lifetimes of the levels in the quadrupole structure have been
 measured. The deduced $B(E2)$ values decrease with increasing spin (and also
 increase of $\it{J}$$^{(2)}$/$B(E2)$ values), indicating the band (structure I) to
 be of AMR origin, followed by a sudden increase at
 spin-parity 63/2$^{+}$. The configuration $\nu{h}_{11/2}^{-2}$
 $\pi({d}_{5/2}/{g}_{7/2})^{-3}$ ${\otimes}$ $ \pi h_{11/2}^{2}$ has been
 assigned to the structure I from the semi-classical model calculation, which
 clearly shows that the quadrupole band (structure I) in $^{143}$Eu originate
 due to the antimagnetic rotation. For structure II a tentative configuration
 of [$\nu{h}_{11/2}^{-2}$ $\pi({d}_{5/2}/{g}_{7/2})^{-3}\pi h_{11/2}^{2}$ + 
$core$(3$\hbar$)]$_{59/2}$ ${\otimes}$ $\pi h_{11/2}^{2}$ has been assigned.
 The $B(E2)$ values for the states in quadrupole structure II have been well
 reproduced within the framework of semi-classical shears model from which 
it may be concluded that this band may also be originated due to AMR, but the possibility of smooth band termination can not be ruled out. The sudden 
rise of the
 $B(E2)$ value at 63/2$^{+}$ may be due to crossing of the two antimagnetic
 rotational band. However such band crossing phenomenon is not incorporated in
 the present semi-classical model \cite{santo, srsc1}. Detailed investigation
 in the framework of microscopic model may lead to a better understanding of
 this behavior. The present measurements conclusively establish the
 antimagnetic rotation in $A$ $\sim$ 140 region. This
 observation outside the $A$ $\sim$ 100 region establishes the antimagnetic
 rotation phenomenon as an alternative mechanism for generation of high
 angular momentum states in weakly deformed nuclei.

\begin{center}
$\textbf{Acknowledgements}$
\end{center}

The authors gratefully acknowledge the financial support by the Department of
 Science $\&$ Technology (DST) for INGA  project (No. IR/S2/PF-03/ 2003-II).
 We would like to acknowledge the help from all INGA collaborators. We are
 thankful to the Pelletron staff for giving us steady and uninterrupted
 $^{31}$P beam. S. R would like to acknowledge the financial assistance from
 the Council of Scientific $\&$ Industrial Research (CSIR), Government of
 India under Contract No. 09/489(0083)/2010-EMR-1. A. B (Contract No.
 09/489(0068)/2009-EMR-1) and S. N (Contract No. 09/081(0704)/2009-EMR-I)
 also would like to acknowledge CSIR for financial support. G. G acknowledges
 the support provided by the University Grants Commission, Departmental
 Research Support (UGC-DRS) Program.


\end{document}